\documentclass{amsart}
\usepackage{amsthm}
\usepackage{amsmath}
\usepackage{mathrsfs}
\usepackage{amssymb}

\theoremstyle{definition}
\newtheorem{definition}{Definition}

\theoremstyle{plain}
\newtheorem{theorem}[definition]{Theorem}
\newtheorem{lemma}[definition]{Lemma}
\newtheorem{corollary}[definition]{Corollary}
\theoremstyle{remark}
\def\vs{}

\title{}

\begin{document}


\noindent
{\Large\sl  You Cannot Press Out the Black Hole}

\vspace{12pt}
\noindent
{\large\sl Daisuke Ida and Takahiro Okamoto}

\vspace{6pt}
\noindent
{\sl Department of Physics, Gakushuin University, Tokyo 171-8588, Japan}\\


\noindent
Abstract.
{\small\it
It is shown that a ball-shaped black hole region homeomorphic with
 $D^n$ cannot be pressed out, along whichever axis penetrating the black
hole region, into a black ring with a doughnut-shaped black hole region
homeomorphic with $S^1\times D^{n-1}$.
A more general prohibition law for the change of the topology of
black holes, including a version of no-bifurcation theorems for black holes,
is given.
}

\vspace{12pt}
\noindent 1. {\it Introduction}
\hspace{12pt}
We would like to discuss here the dynamical aspects of black hole space-times.
The analysis of such nonstationary space-times is often difficult
due to the lack of the geometrical symmetry of the space-time.
However, several results have been known,
which are kinematical, in the sense that they are deduced from 
causal and topological structures of space-times,
but not sensitive to the details of
 the  Einstein equation.

Among these
is on the topology of black hole horizons.
It is known that an apparent horizon in a 4-dimensional space-time
must be diffeomorphic with the 2-sphere~\cite{Haw72,J-V95,B-G95}.
In particular, the stationary black hole event horizon must be the topological 2-sphere.
This follows from the fact that the event horizon coincides with the apparent horizon
in every stationary black hole space-time.
This theorem for the topology of black holes does not exclude an event 
horizon with nonspherical topology.
In fact, for black holes formed by gravitational collapses,
the spatial section of the event horizon can in general be homeomorphic with the torus,
or even with the closed orientable surface of arbitrarily high genus.

An attractive point of view for the topology of the event horizons 
is proposed by Siino~\cite{Sii98}. 
His observation is that the topological information
of the event horizon is encoded in
the crease set of the event horizon, 
which is the set consisting of past end points of 
all the null geodesic generators of the event horizon.
Actually, the crease set of
 the event horizon is quite relevant to
the change of the topology of the black hole.
In particular, the spatial section of the event horizon changes 
its topology only 
when the spatial hypersurface, in which it is embedded,
intersects the crease set of the event horizon.
On the other hand,
Ida~\cite{Ida10} has shown that the crease set has 
the same homotopy type with that of the 
world hypersurface of the event horizon.

Another well-known result, which we would like to mention, 
is the no-bifurcation theorem for black holes~\cite{Haw72,H-E},
 stating that a black hole cannot bifurcate into two or more black holes.
Ida and Siino~\cite{I-S07} observe that there is
a more general form of
the prohibition law for the change of the topology
of black holes than the no-bifurcation law, 
that is, there are many other
forbidden processes in the black hole dynamics.
A typical example of the prohibition law concerns the formation 
of a black ring, by which we mean a black hole with a horizon 
homeomorphic with $S^1\times S^k$ $(k\ge 1)$, 
from a spherical black hole.
There are essentially two kinds of such processes.
Let us consider a black hole in an $n$-dimensional space.
Let the black hole region be a topological
$n$-disk.
Then, the black hole horizon is its boundary $(n-1)$-sphere.
 The black hole region will undergo deformation
 with the time development of the space-time.
Let a pair of horn-shaped black hole regions
 grow  from the body of the black hole, and then
let tips of the horns merge.
Then, we will obtain a black ring, which has
a doughnut-shaped black hole region homeomorphic
with $S^1\times D^{n-1}$.
Another way to get a black ring is by pressing out the ball-shaped
black hole region along whichever axis penetrating the black hole region.
It is argued in Ref.~\cite{I-S07} that the latter 
``pressing-out process'', which corresponds to the 
``white $(n-2)$-handle attachment'' 
in the terminology of Ref.~\cite{I-S07},
cannot be realized in any space-time. 
However, the reasoning in Ref.~\cite{I-S07} requires a technical 
assumption such that 
the time function induced on a slightly deformed event horizon 
is a Morse function on it, which is not fully satisfactory.

In the rest of the paper, 
we show, under milder conditions, 
that a spherical black hole cannot be pressed out into a black ring.

\vspace{12pt}
\noindent 2. {\it Preliminaries}
\hspace{12pt}
We mean by a space-time $M$ a Hausdorff, orientable, time-orientable, Lorentzian manifold\footnote{By Lorentzian 
manifold, we mean a differentiable manifold
endowed with a nondegenerate, smooth metric tensor field with a signature
of $(-,+,+,\cdots,+)$.
} with ${\rm dim}M\ge 4$.
Note that the existence of a Lorentzian metric 
guarantees the paracompactness.
Here, we consider weakly asymptotically simple  and strongly future asymptotically predictable space-times~\cite{H-E}.
The former condition is one of the standard conditions for the space-time
to be asymptotically flat, and the latter requires that
there is a partial Cauchy surface $S$ such that $\mathscr{I}^+\subset \overline{D^+(S)}$ and $\overline{J^-(\mathscr{I}^+)}\cap J^+(S)\subset D^+(S)$
hold.

If the space-time $M$ is strongly future asymptotically predictable from a partial Cauchy surface $S$, 
there is a homeomorphism 
\begin{eqnarray*}
  f: [0,\infty)\times S\to D^+(S);(\tau,x)\mapsto S_\tau(x),
\end{eqnarray*}
such that the conditions
\begin{enumerate}
\item for each $\tau \in [0,\infty)$, 
$S_\tau=f(\tau,S)$ is a partial Cauchy surface homeomorphic with $S$,

\item $S_0=S$,
\item for $0<\tau<\tau'$,
$S_{\tau'} \subset I^{+}(S_\tau)$,

\item for $0 < \tau < \tau '$ and for
each $p\in S_\tau$, 
every future inextendible causal curve starting from $p$
intersects $S_{\tau '}$ or $\mathscr{I}^{+}$,
\end{enumerate}
hold (See Ref.~\cite{H-E} Prop.~9.2.3.).

The function $t=p_1\circ f^{-1}: D^+(S)\to [0,\infty)$ can be regarded as
a time function on $D^+(S)$, where $p_1$ denotes the projection onto the
first component.
Thus, due to the existence of the global time coordinate $t$, 
the black hole region at time: $t=\tau$ can be naturally 
defined.

\vs
\begin{definition}[Black hole region]
For each $\tau\in [0,\infty)$,
the black hole region $B_\tau$ at time $t=\tau$ 
is defined by
\begin{eqnarray*}
  B_\tau=S_\tau\cap(M\setminus\overline{J^-(\mathscr{I}^+)}).
\end{eqnarray*}
\end{definition}
\vs

In the following, we assume that each partial Cauchy surface $S_\tau$
is a smooth hypersurface in the space-time $M$, and the time function $t$
is a smooth function on $D^+(S)$.

\vspace{12pt}
\noindent
3. {\it ``Black holes are not pressed out'' theorem}
\hspace{12pt}
In order to formulate whether or not the black hole region is pressed out,
we focus on the topological property of simple closed curves in the
black hole region.

At some initial time,
let a simple closed curve $L$ in the black hole region
have the property that it bounds a 2-disk in the black hole region.
If $L$ lost this property after a while, we could say that
the black hole region is pressed out.
Conversely, there would be such a simple closed curve with this property, 
whenever the black hole region is pressed out.

We formulate this by introducing a couple of definitions.

\vs
\begin{definition}[B-contractibility]
A simple closed curve $L:S^1\to B_\tau$, which is a smooth embedding in 
$B_\tau$, is said to be B-contractible
in $B_\tau$, 
if it is contractible to a point within $B_\tau$, 
that is  if there is a homotopy 
$f:S^1 \times [0,1] \rightarrow B_\tau$ 
such that 
the conditions:
\begin{enumerate}
\item for each $\theta\in S^1$, $f(\theta,0)=L(\theta)$ holds,
\item $f(~\cdot,1):S^1\to B_\tau$ is a constant map, 
\end{enumerate}
are satisfied.
\end{definition}
\vs

\vs
\begin{definition}[Descendent of a simple closed curve]
  Let $L:S^1\to S_{\tau_i}$ ($\tau_i\ge 0$) be a simple closed curve
smoothly embedded in $S_{\tau_i}$.
For $\tau_f > \tau_i$, a simple closed curve $L':S^1\to S_{\tau_f}$
is called a descendent of $L$, if there is a compact
Lorentzian 2-submanifold $N$ smoothly embedded in $D^+(S_{\tau_i})$,
whose boundary is the disjoint union $L(S^1)\sqcup L'(S^1)$.
\end{definition}
\vs

In this terminology, we state our main theorem claiming that any black hole cannot
be pressed out.

\vs
\begin{theorem}\label{thm:main}
Every descendent of 
a B-contractible simple closed curve 
is B-contractible.
\end{theorem}
\vs

In order to prove this theorem, we need the following lemma
asserting that a nonsingular timelike vector field on
a closed Lorentzian submanifold of the space-time $M$ can be
globally extended to that on $M$.

\vs
\begin{lemma}\label{lem:vectX}
For any closed Lorentzian submanifold $N$ smoothly embedded in 
the space-time $M$,
there exists a smooth, future-directed, timelike vector field in $M$,
which is tangent to $N$, everywhere on $N$.
\end{lemma}
\vs

\begin{proof}
For any point $p\in N$, there is an open neighborhood $U_p$ of $p$ 
such that $\overline U_p$ is compact and that 
there is a nonsingular future-directed timelike
 vector field $Y_p$ on $U_p$, which is tangent to $N$ 
everywhere on $N\cap U_p$,
for the space-time $M$ 
is a time-oriented Lorentzian manifold.
Let $\{V_\alpha\}_{\alpha\in A}$ be an open cover of $M\setminus N$, 
such that
each $\overline V_\alpha$ is compact and that
on each $V_\alpha$, there is a nonsingular future-directed 
timelike vector field $Y_\alpha$.
Since $\{U_p\}_{p\in N}$, $\{V_\alpha\}_{\alpha\in A}$ is 
an open cover of $M$, and $M$ is a paracompact manifold, 
there is a locally finite refinement $\{W_j\}_{j\in J}$ of 
$\{U_p\}_{p\in N}$, $\{V_\alpha\}_{\alpha\in A}$.
Then, for each $j\in J$, $\overline W_j$ is compact.

For each $j\in J$ such that $W_j\cap N\ne \emptyset$ holds, 
we can choose a point $p\in W_j\cap N$ such that
$W_j$ is contained in $U_p$.
Then, define a local vector field $X_j$ on $W_j$ as the 
restriction of $Y_p$ on $W_j$. 

On the other hand, for each $j\in J$ such that
$W_j\cap N=\emptyset$ holds, 
$W_j$ is contained in $V_\alpha$ for some $\alpha\in A$,
or otherwise in $U_p$ for some $p\in N$. 
Then, define a local vector field $X_j$ on $W_j$ as the restriction 
of either $Y_\alpha$ or $Y_p$ on $W_j$.

There is a partition of unity subordinate to the 
cover $\{W_j\}_{j\in J}$, that is,
there is a collection of smooth functions $\{\rho_j\}_{j\in J}$ 
on $M$ such that 
\begin{enumerate}
\item for each $j\in J$, $0\le \rho_j\le 1$ on $M$, 
\item for each $j\in J$, ${\rm supp}(\rho_j)\subset W_j$,
\item for every point $p\in M$, $\sum_j \rho_j(p)=1$
\end{enumerate}
hold.

Define the vector field $X$ on $M$ by
\begin{eqnarray*}
  X=\sum_{j\in J}\rho_j X_j.
\end{eqnarray*}
Since $\{W_j\}_{j\in J}$ is locally finite, every point in $M$
has a neighborhood such that $\sum_j\rho_jX_j$ is a finite sum.

Clearly, $X$ is a nonsingular, future-directed, timelike vector field
on $M$, which is tangent to $N$, everywhere on $N$.
This completes the proof of  Lemma~\ref{lem:vectX}.
\end{proof}

Now, we are in a position to prove  Theorem~\ref{thm:main}.
\begin{proof}[Proof of Theorem~\ref{thm:main}]
Let $L:S^1\to B_{\tau_i}$ be a B-contractible simple closed curve in the black hole region $B_{\tau_i}$.
Then, there is a homotopy $f:S^1\times [0,1]\to B_{\tau_i}$ between $L$ and the constant map: $S^1\to \{p\}$, $p\in B_{\tau_i}$.

For $\tau_f>\tau_i$, let $B_{[\tau_i,\tau_f]}$ 
be the portion of 
the black hole region
$M\setminus \overline{J^-(\mathscr{I}^+)}$ 
between $B_{\tau_i}$ and
 $B_{\tau_f}$: 
$B_{[\tau_i,\tau_f]}=
\{x\in M\setminus \overline{J^-(\mathscr{I}^+)}; t(x)\in [\tau_i,\tau_f]\}$.
Let $N$ be a smoothly embedded 2-dimensional compact Lorentzian submanifold of 
$B_{[\tau_i,\tau_f]}$, whose boundary consists of 
$L(S^1)$ and a simple closed curve $L'(S^1)$
in $S_{\tau_f}$. 
$N$ is properly embedded in $B_{[\tau_i,\tau_f]}$.
By Lemma~\ref{lem:vectX}, there is a nonsingular smooth timelike vector field $X$ on
$B_{[\tau_i,\tau_f]}$, tangent to $N$ everywhere on $N$.
The vector field $X$ gives a diffeomorphism
$\varphi: B_{\tau_i}\times [0,1]\to  B_{[\tau_i,\tau_f]}$,
that is,
for each $(x,r)\in B_{\tau_i}\times [0,1]$, 
$\varphi(x,r)$ is defined to be the point where the integral curve
of $X$ starting from $x$ intersects 
$B_{(1-r)\tau_i+r\tau_f}$.
Then, the descendent $L'$ of $L$ can be
written as
\begin{eqnarray*}
L':S^1\to B_{\tau_f};\theta\mapsto\varphi(L(\theta),1).
\end{eqnarray*}

Consider now the continuous map
\begin{eqnarray*}
f':S^1\times [0,1] \to B_{\tau_f};
(\theta,s)\mapsto \varphi(f(\theta,s),1).
\end{eqnarray*}
For $f'(\theta,0)=L'(\theta)$ and 
$f'(\theta,1)=\varphi(p,1)$,
the map $f'$ gives a homotopy 
within closed curves in $B_{\tau_f}$,
between $L'$ and the constant map onto $\varphi(p,1)$.
Thus, the descendent $L'$ of $L$ is B-contractible.
This completes the proof of Theorem~\ref{thm:main}.
\end{proof}

\vspace{12pt}
\noindent 
4. {\it Extension for B-contractible $k$-spheres}
\hspace{12pt}
Theorem~\ref{thm:main} can be easily 
extended for ``B-contractible spheres''
with general dimensions.
We briefly state the extended theorem here.
First, we need a few definitions.

\vs
\begin{definition}[B-contractible $k$-sphere]
For $0\le k\le {\rm dim}M-2$,  let $L:S^k\to B_{\tau}$
be a smooth embedding of a $k$-sphere into the black hole 
region at $t=\tau$. (The $0$-sphere is a pair of points).
Then, $L$ is said to be B-contractible in $B_\tau$,
if it is contractible to a point within $B_\tau$.
\end{definition}
\vs

Accordingly, the notion of a descendent of an embedded $k$-sphere
is defined as follows.

\vs
\begin{definition}[Descendent of an embedded $k$-sphere]
For $0\le k\le {\rm dim}M-2$,   let $L:S^k\to S_{\tau_i}$ ($\tau_i\ge 0$)
be a smooth embedding. For $\tau_f>\tau_i$,
a smooth embedding $L':S^k\to S_{\tau_f}$ is said to be
a descendent of $L$, if there is a compact Lorentzian $(k+1)$-submanifold
 smoothly embedded in $D^+(S_{\tau_i})$, whose boundary is the disjoint
union $L(S^k)\sqcup L'(S^k)$.
\end{definition}
\vs

Then, the generalization of Theorem~\ref{thm:main} is
stated as follows.

\vs
\begin{theorem}\label{thm:main2}
  Every descendent of a B-contractible $k$-sphere is B-contractible.
\end{theorem}
\vs

We omit the proof, for it parallels to that of Theorem~\ref{thm:main}.
Theorem~\ref{thm:main2} reduces, in the case of $k=0$, to a version of 
``no-bifurcation theorems''
 implying that any black hole cannot bifurcate
into several black holes.

\vs
\begin{corollary}[No-bifurcation theorem for black holes]
Let $p$ and $q$ be a pair of points belonging to the same connected
component of the black hole region $B_{\tau_i}$.
For $\tau_f>\tau_i$,
if  $p'$ belongs to $I^+(p)\cap B_{\tau_f}$
and $q'$ belongs to $I^+(q)\cap B_{\tau_f}$,
then $p'$ and $q'$ belong to the same connected component
of the black hole region $B_{\tau_f}$.
\end{corollary}
\vs

\begin{proof}
This can be seen by noting that $L:S^0\to B_{\tau}$ is B-contractible,
if and only if $L(S^0)$ belongs
to the same connected component of $B_{\tau}$.  
\end{proof}

\vspace{12pt}
\noindent
5. {\it Concluding remarks}
\hspace{12pt}
We have defined the notion of the ``pressing-out process''
for black hole event horizons in terms of 
the ``B-contractiblity'' of 
simple closed curves
interpolated by a Lorentzian cobordism.
Then, we have shown that these pressing-out processes 
are never realized (Theorem~\ref{thm:main}). 
One consequence derived from this is that an axisymmetric black hole
cannot be converted into a black ring in any axisymmetric 
process.
This can be seen as follows.
Let $B_{\tau_i}\simeq D^n$ (homeomorphism) be a ball-shaped 
black hole region 
in an $(n+1)$-dimensional axisymmetric
space-time $M$,
and it is converted into a doughnut-shaped black hole region
$B_{\tau_f}\simeq S^1\times D^{n-1}$ after a while,
such that each orbit of the $U(1)$-isometry on a point in $B_{\tau_f}$
is an incontractible circle in $B_{\tau_f}$.
There will be a timelike curve $\gamma:[0,1]\to M$ 
connecting a point $\gamma(0)\in B_{\tau_i}$
and a point $\gamma(1)\in B_{\tau_f}$.
By the standard arguments of general position, we can assume
that the timelike curve $\gamma$ is smooth and 
it does not contain a fixed point
of the $U(1)$-isometric action.
Then, the orbit of $U(1)$-isometry on $\gamma$ will be a
Lorentzian 2-submanifold smoothly embedded in $M$, which interpolates
a loop $L$ in $B_{\tau_i}$ and 
a loop $L'$ in $B_{\tau_f}$.
The loop $L$ is a B-contracible simple closed curve in $B_{\tau_i}$,
for $B_{\tau_i}$ is simply connected,
while the loop $L'$ is not B-contractible, for it is an orbit of the
$U(1)$-isometry on $\gamma(1)\in B_{\tau_f}$.
This is impossible by Theorem~\ref{thm:main}.
In this way, the formation of a black ring from a Kerr 
(or Myers-Perry) black hole
must involve a non-axisymmetric process.

Theorem~\ref{thm:main2} controls more complicated dynamical evolutions
of the topology of black holes in higher-dimensional space-times.
It is known that the black hole no-hair property does not in general
 hold in higher-dimensional space-times. 
Rather,  various types of black hole with nontrivial horizon 
topologies are allowed by the Einstein equation.
These black holes are not necessarily stable in dynamics\cite{G-L93}.
This implies that 
various dynamical processes involving the 
topology change of the black hole event horizon naturally take place.
Our result here would serve as a good starting point to 
qualitatively  understand such dynamically complicated processes.

Let us finally comment on the relationship between 
the previous work~\cite{I-S07} and the present one.
In Ref.~\cite{I-S07}, the dynamics of the topology of the event horizon
is described in terms of a Morse function defined on an appropriately
smoothed event horizon as a world hypersurface.
Here, each topology change of the black hole region corresponds to
a critical point of the Morse function and it is therefore
classified using the Morse index $\lambda$ of the critical point,
where $\lambda=0,\cdots,n$.
Furthermore, each critical point is classified a priori
into black or white according to the relative orientation 
between the black hole region and the external region around the
critical point.
Hence, each topology change is identified geometrically
 with black or white $\lambda$-handle attachment.
In this way, 
any complicated topological dynamics of the black hole reduces to 
a comprehensible local dynamics.
Then, it is argued in Ref.~\cite{I-S07} that 
any change of the horizon topology corresponding to the
white $\lambda$-handle attachment never 
occurs.
This reasoning relies entirely on the existence of the Morse function
on the event horizon, which is not guaranteed in general.
The present work overcomes this drawback in terms of a global approach,
which does not need such Morse functions.
Theorem~\ref{thm:main2} for 
the B-contractible $k$-sphere is
essentially the same as the statement that excludes the
white $(n-k-1)$-handle attachment in the terminology of the
local approach in Ref.~\cite{I-S07}.

\end{document}